\documentclass[aps,preprint,showpacs,epsfig,epsf]{revtex4}

\usepackage{graphicx}
\usepackage{amssymb,amsbsy}
\usepackage{amsmath}

\begin{document}

\title{Scattering Properties of Paramagnetic Ground States in the Three-Dimensional Random-Field Ising Model}

\author{Gaurav P. Shrivastav$^1$, Siddharth Krishnamoorthy$^{2}$,  Varsha Banerjee$^2$ and Sanjay Puri$^1$\\
$^{1}$School of Physical Sciences, Jawaharlal Nehru University,
New Delhi -- 110067, India.\\
$^2$Department of Physics, Indian Institute of Technology,\\ Hauz Khas, New Delhi -- 110016, India.\\ }

\begin{abstract}
We study the ground-state ($T = 0$) morphologies in the $d = 3$  random-field Ising model (RFIM) using a computationally efficient graph-cut method. We focus on paramagnetic states which arise for disorder strengths $\Delta > \Delta_{c}$, where $\Delta_{c}$ is the critical disorder strength at $T = 0$. These paramagnetic states consist of correlated ``domains'' of up and down spins which are separated by rough, fractal interfaces. They show novel scattering properties with a cusp singularity in the correlation function at short distances. 
\end{abstract}


\maketitle

Spin systems with quenched disorder have challenged physicists for several decades. The competing nature of the interactions creates difficulties in studying them analytically and computationally. As a result,  the properties of phases and phase transitions in disordered systems remain controversial. The {\it random-field Ising model} (RFIM) is an archetypal example of a system with quenched disorder and is described by the Hamiltonian \cite{nv,ntrmn}: 
\begin{equation}
\label{rfim}
E = -J\sum_{\langle ij \rangle} \sigma_i \sigma_j - \sum_{i=1}^{N} h_i \sigma_i, \quad \sigma_{i}=\pm 1.
\end{equation}
Here,  $J>0$ is the strength of the exchange interaction between nearest-neighbor spins. The variables $\{h_{i}\}$ are random fields, usually drawn from a Gaussian distribution whose standard deviation $\Delta$  is a measure of  disorder.  The phase diagram of the RFIM has been the subject of much discussion. In the 2-dimensional case ($d = 2$), there is no {\t long-range order}  in the presence of disorder, no matter how small. However, in $d = 3$, there is a small region of ($T,\Delta$)-values where the equilibrium phase is ferromagnetic \cite{ji84,bk87}.  Let us focus on the case with zero temperature ($T = 0$). In that case, the system exhibits a phase transition from a ferromagnetic phase (for $\Delta < \Delta_c$) to the paramagnetic  phase (for $\Delta > \Delta_c$). The nature of this transition has received considerable attention \cite{my92,hr95,aas97}. An important  study of the $d=3$ RFIM is due to Middleton and Fisher \cite{mf}. They studied a wide range of physical properties and convincingly demonstrated that there is a second-order phase transition at $\Delta = \Delta_c$. 

At $T = 0$, all the information about the system is encoded in the ground-state. Further, according to the {\it zero-temperature fixed point hypothesis}, transitions at $T = 0$ and $T \ne 0$ are in the same universality class  \cite{villain,fisher}. Therefore, a study of the ground-state morphology is important in understanding the RFIM phase diagram in $d = 3$. A typical method of accessing the ground-state is via Monte Carlo (MC) evolution (e.g., Metropolis \cite{met}, Simulated annealing \cite{sa}, etc.) from an arbitrary initial condition. However, MC approaches for disordered systems suffer from several drawbacks. First, the competition between exchange interactions and the random field introduces deep valleys in the free-energy landscape. These metastable states trap the evolving system and  impede the relaxation to the ground-state. The system then opts for a {\it local minimum}, which can be far removed from the {\it global minimum}, and may not reflect any of its properties. Further, as the MC techniques involve $\sim$ order (1) spin-flip at a time, the possibility of escape from a local minimum to the  global minimum is small. Second, MC methods suffer from a non-polynomial (NP) divergence of computation time with system size. Thus, it is computationally very demanding to reach the global minimum for large systems with disorder. Consequently we still do not have a complete understanding of the nature of the ground-state. 

To address this problem, several optimization techniques based on ``max-flow/min-cut'' or ``graph-cuts'' have been developed for a wide class of energy functions (or Hamiltonians) of binary variables \cite{ff,gt,bk}. The basic approach in a {\it graph-cut method} (GCM) is to construct a specialized graph for the energy function to be minimized such that the minimum cut on the graph also minimizes the energy. The cut  enables simultaneous relabeling of several spin variables or nodes. As a consequence, an exponentially large portion of the phase space can be sampled in a single move, thereby facilitating a quick search for a global minimum or a ``good-quality'' local minimum. Typically, the search time in these procedures has a polynomial dependence on the system size. An important class of energy functions are those which are  (a) quadratic and (b) satisfy a ``regularity'' condition. In that case, the max-flow/min-cut technique actually yields the global minimum or exact ground state of the energy function in polynomial time \cite{ps1982,kz,pr75}. The Hamiltonian of the RFIM specified  in Eq.~(\ref{rfim}) belongs to this class \cite{auriac}. We are therefore assured of reaching the exact ground state of the RFIM if energy minimization is via graph-cuts. 

The literature on combinatorial optimization provides many graph-cut algorithms with different polynomial complexity times. Some of the standard approaches include the Ford-Fulkerson (FF) method of augmenting paths \cite{ff}, the Goldberg-Tarjan (GT) push-relabel method \cite{gt}, and  the more recent Boykov-Kolmogorov (BK) method \cite{bk}. A benchmarking of the above algorithms on a number of typical graphs has revealed that the BK method works several times faster than any other. While the FF and GT algorithms exhibit an $N^3$-dependence on the system size $N$, the BK method is linear in $N$ \cite{bk}. 

In this paper, we use the BK method to study the ground-state ($T$=0) morphologies in the $d = 3$ RFIM. This enables us to access exact ground states for substantially larger system sizes than in previous studies \cite{mf}. We need these large sizes to obtain smooth data for statistical properties of the morphology, e.g., correlation function, structure factor, etc. We focus on the scattering properties of the domain structure in the paramagnetic state, i.e., for field strengths $\Delta > \Delta_{c}$. This domain morphology has several non-trivial features, which we highlight in this paper. The correlation function $C\left(r,\Delta \right) (=\langle \sigma_{i} \sigma_{j} \rangle - \langle\sigma_{i} \rangle \langle \sigma_{j} \rangle$ with $r = \vert \vec{r_i} - \vec{r_j} \vert $) is a scaling function of $r/\xi\left(\Delta\right)$, where the correlation length $\xi$ diverges as $\Delta \rightarrow \Delta_{c}^{+}$. At small values of  $r/\xi$, $C\left(r,\Delta\right)$ exhibits a cusp singularity characterized by the roughness exponent $\alpha$:  $C\left(r,\Delta\right) \simeq 1 - A\left(r/\xi\right)^{\alpha} + \cdot\cdot\cdot$. This singularity has important consequences for the high-momentum behavior (``tail'') of the structure factor. A similar cusp has been reported earlier in the context of {\it fluctuation-dominated phase separation} \cite{db,mr}, and is a consequence of soft and ragged interfaces separating equilibrium phases. In the paramagnetic phase of the RFIM, there are no coexisting equilibrium phases. Nevertheless, there exist correlated domains of size $\sim \xi$ which are enriched in up or down spins. The scattering properties of these domain boundaries are analogous to those of fractal interfaces. We also provide accurate estimates of the critical point $\Delta_{c}$ and the correlation length exponent $\nu$ calculated from the ground-state morphologies.

Before presenting our results, we discuss the graph-cut approach, which has many potential applications for energy minimization in complex spin systems. This method can be applied to energy functions of the form:  
\begin{equation}
\label{cfn}
E(\{s_{i}\}) =  \sum_{\{ij\} \in \mathcal{N}}V_{ij}(s_i,s_j) + \sum_{i\in \mathcal{S}}D_i(s_i).
\end{equation}
The label $s_{i}$ of site $i$ $\in \mathcal{S}$ can take a value 0 or 1, and the sites are related to one another by a well-defined neighborhood $\mathcal{N}$. The function $D_{i}$ measures the cost of assigning the label $s_{i}$ to the site $i$, and $V_{ij}(s_i,s_j)$ measures the penalty (or cost) of assigning labels $s_{i}$ and $s_{j}$ to adjacent sites $i$ and $j$. 

The starting point in a GCM is to construct a specialized graph for the energy function $E$ such that the minimum cut on the graph  yields minimization of the energy.  A graph $\mathcal{G}$ is an ordered pair of disjoint sets $(\mathcal{V},\mathcal{E})$, where  $\mathcal{V}$ is the set of vertices and $\mathcal{E}$ is the set of edges. An edge  $ij$  joining vertices $i$ and $j$  is assigned a weight $V_{ij}$.  A cut $C$ is a partition of the vertices $\mathcal{V}$ into two sets $\mathcal{R}$ and $\mathcal{Q}$. Any edge ${ij}$ $\in$ $\mathcal{E}$ with $i$ $\in$ $\mathcal{R}$ and $j$ $\in$ $\mathcal{Q}$ (or vice-versa) is a cut edge. The cost of the cut is defined to be the sum of the weights of the edges crossing the cut. The minimum-cut problem is to find the cut with the smallest cost. 

The energy function $E$ must satisfy the regularity condition for it to be graph-representable.  Regularity is defined by the inequality $V_{ij}(0,0) + V_{ij}(1,1) \leq V_{ij}(1,0)+V_{ij}(0,1)$ \cite{kz}. The spin variables ($\sigma_{i} = \pm 1$) in Eq.~(\ref{rfim}) can be transformed into occupation-number variables ($n_{i} = 0, 1$) through the transformation $n_{i} = (1+\sigma_{i})/2$. Then, neglecting constant terms,
\begin{equation}
\label{rfimo}
E(\{n_{i}\}) =  -4J\sum_{\langle ij \rangle} n_i n_j  -2 \sum_{i = 1}^{N} \left(h_{i} - qJ\right) n_i, \quad n_{i} = 0,1,
\end{equation}
where $q$ denotes the number of nearest neighbors of a lattice site. It is straightforward to check that the interaction term of Eq.~(\ref{rfimo}) satisfies the regularity condition. Thus, the energy function $E$ of Eq.~(\ref{rfim}) is graph-representable and can be minimized using a GCM to yield the exact ground state.   
Each iteration of the GCM finds an optimal subset of nodes with a fixed label $s_{i}$  ($=$ 0  or 1) that gives the largest decrease in energy.  This computation is done via graph-cuts on the specialized graph representing the energy function $E$. The algorithm repeatedly cycles through the labels $s_{i}$ until the global minimum is reached. 

Our simulations of the $T=0$ RFIM have been performed on $d=3$ lattices of size $L^{3}$ ($L \le 256$), with periodic boundary conditions applied in all directions. All the statistical data presented here, unless otherwise specified, is for a cubic system with $L = 256$. The initial configuration of the lattice is chosen to be a  random mix of $\sigma_{i} = \pm 1$, corresponding to the paramagnetic state at $\Delta = \infty$ with $\xi = 0$.  The results have been averaged over 100 sets of $\{h_{i}\}$ for each value of  $\Delta$. Our studies indicate that the GCM has a 99\% overlap with the ground-state in the first iteration itself, provided the disorder strength is not too close to the critical value $\Delta_{c}$. (We do observe ``critical slowing down'' in  the GCM as $\Delta \rightarrow \Delta_{c}$, but the phenomenon is much milder than in conventional MC methods.) We also find that the average energy per spin in the ground-state is an order of magnitude less than that obtained by the Metropolis algorithm. As mentioned earlier, the MC evolution invariably gets trapped in high-energy metastable states. 

\begin{center}
\begin{figure}
\includegraphics[width=15cm]{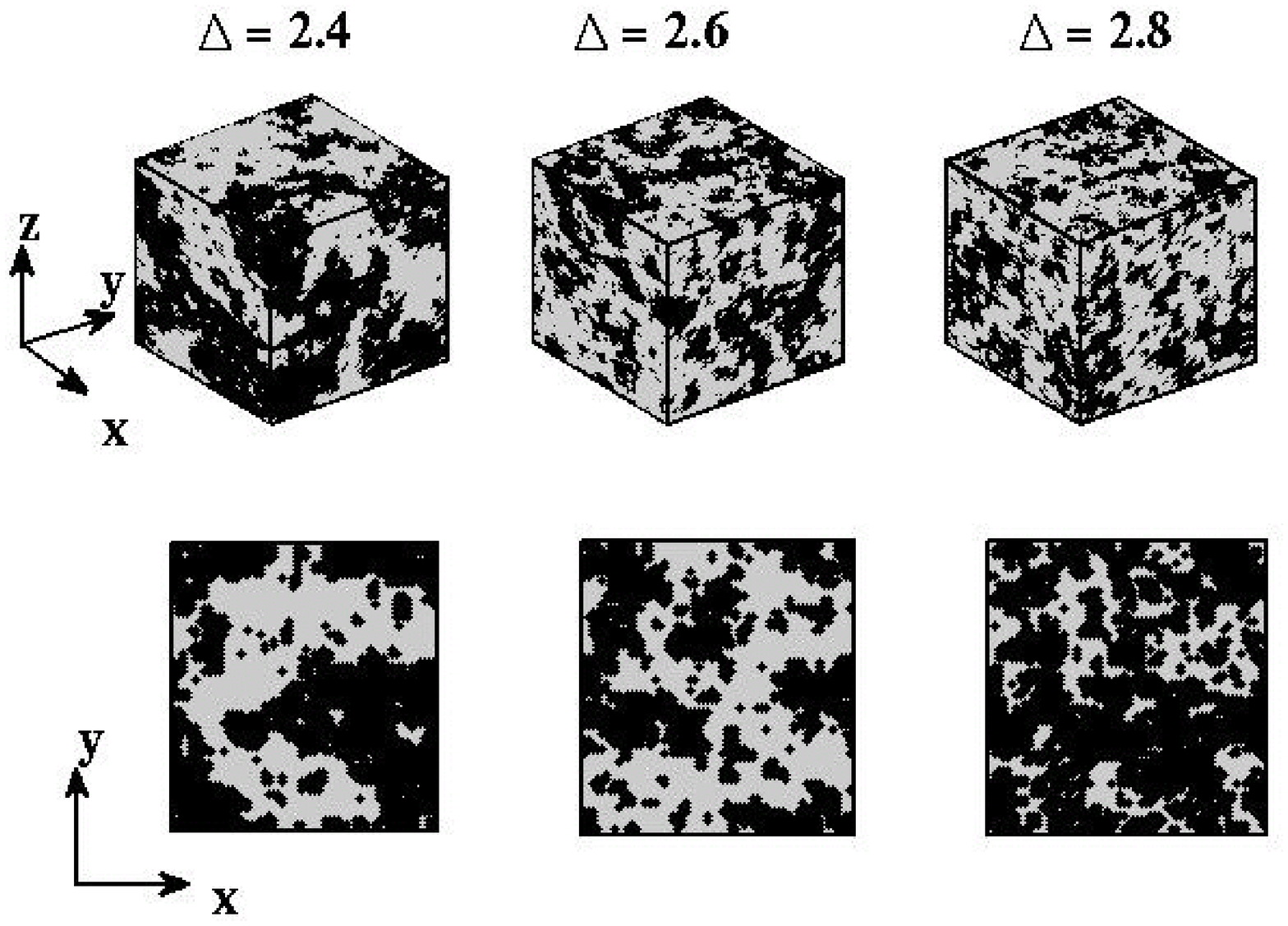}
\caption{Ground-state morphologies of the RFIM obtained using the $\alpha$-expansion GCM, for disorder strengths $\Delta$ $=$ 2.4, 2.6 and 2.8. The snapshots in the top frames correspond to a $ 64^3 $ lattice with periodic boundary conditions in all directions. Regions with up spins and down spins are marked black and grey. The domains shrink in size and interfaces roughen with increasing disorder, as is evident from the cross-sections (taken at $z = 32$) in the bottom frames.}
\label{RFIM}
\end{figure}
\end{center}

\begin{center}
\begin{figure}
\includegraphics[width=15cm]{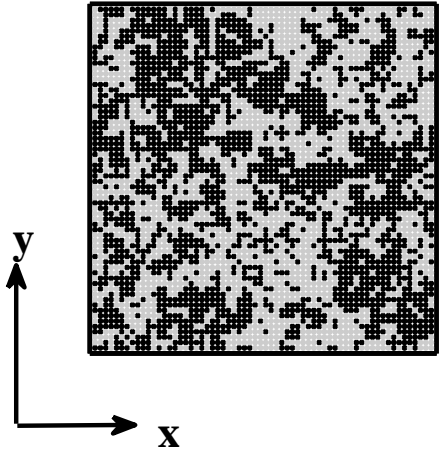}
\caption{Equilibrium morphology of the disorder-free Ising paramagnet for $T = 4.515$ $\left(T_{c}\simeq 4.5103 \ \mbox{for} \ \Delta = 0\right)$. This snapshot is obtained from Glauber MC simulations of a $ 64^3 $ lattice with periodic boundary conditions. The system is evolved from an arbitrary initial condition to its equilibrium state, where the morphology is invariant with time. The above snapshot corresponds to a cross-section (at $z = 32$) of a $64^3$ lattice.}
\label{MC}
\end{figure}
\end{center}

The ground-state morphology in the paramagnetic state has the following features. As the disorder strength is reduced from $\Delta = \infty$, there is emergence of correlated regions or domains of size $\xi$,  enriched in either up or down spins (see snapshots in Fig. \ref{RFIM} for $\Delta$ = 2.4, 2.6, 2.8). These regions grow in size with $\xi \rightarrow \infty$ as $\Delta \rightarrow \Delta_{c}^{+}$. [A similar divergence of $\xi$ is seen for $\Delta = 0$, $ T\rightarrow T_{c}^{+}$. However, in that case, a typical MC snapshot (see Fig. \ref{MC}) shows that the domain morphology is not as compact or well-defined as in Fig. \ref{RFIM}. Fig. \ref{MC} corresponds to $T=4.515$, which is very close to the critical point: $T_{c}=4.5103$ \cite{mf67}.] By computing the Binder cumulant \cite{bh02} as a function of $\Delta$ for different system sizes,   we estimate $\Delta_{c}$ ($T=0$) $\simeq 2.278 \pm 0.002$. For $\Delta < \Delta_{c}$, the ground-state morphology consists of a single large domain of ordered spins (up or down) with small clusters of oppositely-directed spins. The fraction of oppositely-directed spins decreases as $\Delta \rightarrow 0$.

\begin{center}
\begin{figure}
\includegraphics[width=15cm]{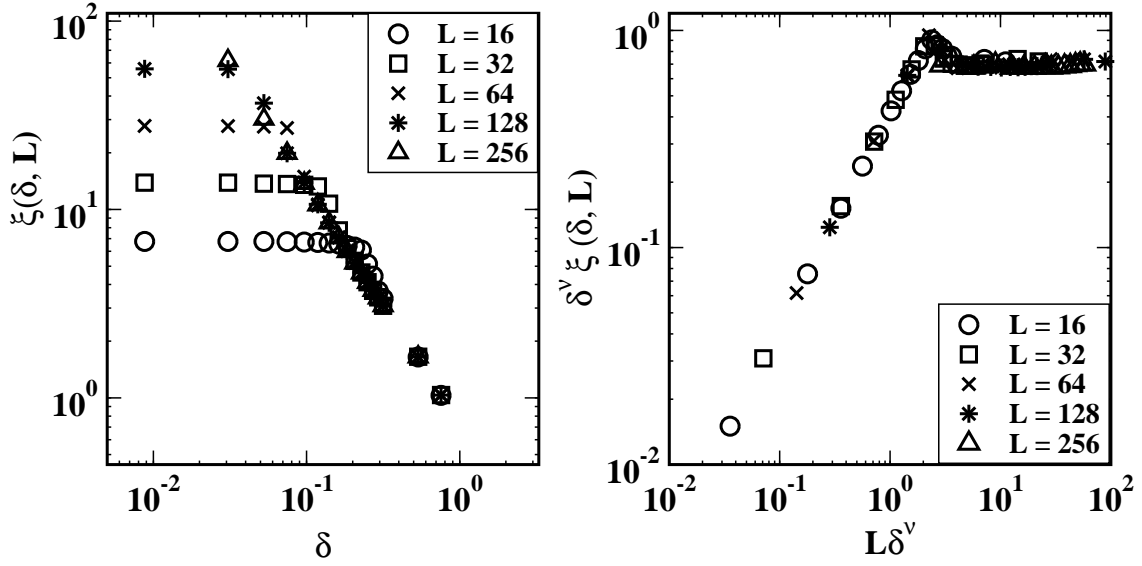}
\caption{(a) Plot of correlation length $\xi(\delta,L)$ vs. $\delta$, where $\delta = \left(\Delta - \Delta_{c}\right)/\Delta_{c}$. We present data for cubic lattices of size $L = $ 16, 32, 64, 128, 256 - denoted by the specified symbols. The data sets were averaged over 100 random-field configurations. The correlation length is defined as the distance over which the correlation function falls to $0.2\times$ maximum value. (b) Data collapse resulting from the finite-size scaling ansatz $\xi(\delta,L) = \delta^{-\nu} f(L \delta^{\nu})$,  yielding the correlation length exponent $\nu \simeq 1.308$.}
\label{FSS}
\end{figure}
\end{center}

What are the quantitative properties of the domain morphologies in Fig. \ref{RFIM}? The standard probe for quantifying these patterns is the correlation function $C\left( r,\Delta\right)$.  The correlation length $\xi(\Delta)$ is defined as the distance over which $C\left(r,\Delta\right)$ decays to (say) $0.2 \times$ maximum value. In Fig. \ref{FSS}(a)  we plot $\xi(\delta,L)$ vs. $\delta$ [where $\delta=(\Delta-\Delta_c)/\Delta_c$] for system sizes $L$ ranging from 16 to 256. The correlation length diverges as $\xi \sim \delta^{-\nu}$ when $\delta \rightarrow 0^{+}$, but this is  limited by the lattice size $L$. The presence of  these finite-size effects  can be used to estimate the critical exponent $\nu$.  The finite-size scaling ansatz  $\xi(\delta,L) = \delta^{-\nu} f(L \delta^{\nu})$ results in the  data collapse seen in Fig. \ref{FSS}(b), yielding $\nu \simeq 1.308\pm0.005$. This is consistent with the earlier results of $\nu = 1.4 \pm 0.2$ (Rieger and Young \cite{ry93}) and $\nu = 1.37 \pm 0.09$ (Middleton and Fisher \cite{mf}).

\begin{center}
\begin{figure}
\includegraphics[width=15cm]{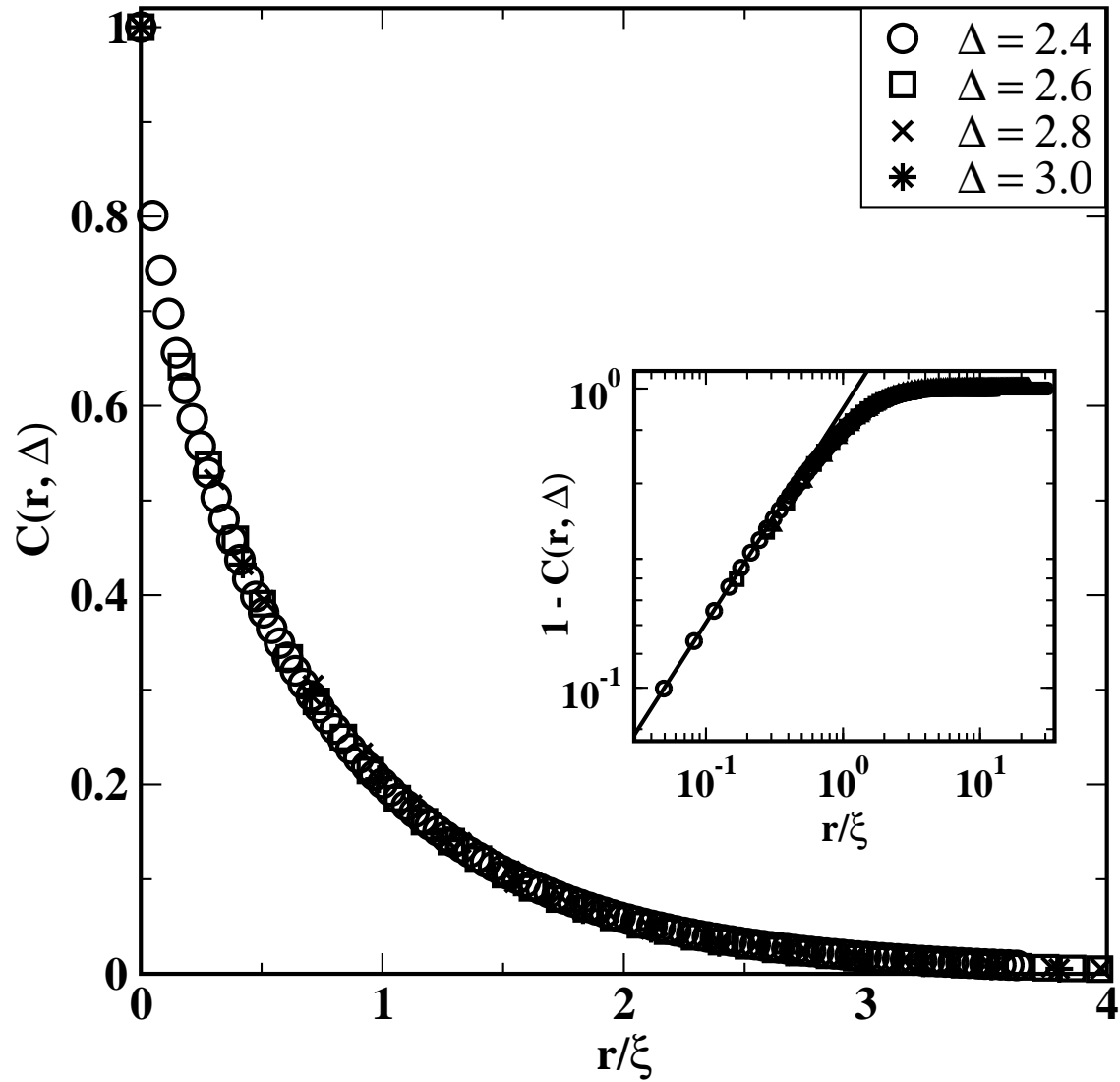}
\caption{Scaled correlation function $\left[C\left(r,\Delta \right) \ \mbox{vs.} \ r/\xi\right]$ for specified disorder strengths. The numerical data has been averaged over 100 random-field configurations for a lattice of size $ 256^3 $. The  inset  shows the small-$r/\xi$ behavior on a log-log scale to highlight the cusp singularity. The slope of the solid line yields the cusp exponent $\alpha \simeq  0.5$.}
\label{SCF}
\end{figure}
\end{center}

If the system is characterized by a single length scale, the morphology of the domains does not change with $\Delta$, apart from a scale factor. In that case, the correlation function exhibits scaling:  $C\left(r,\Delta\right) = g\left(r/\xi\right)$ \cite{pw09}.  This is verified in Fig. \ref{SCF}, where  we plot $C\left( r,\Delta\right)$ vs. $r/\xi$ for different  disorder amplitudes $\Delta > \Delta_{c}$. The data collapse for different values of $\Delta$ is excellent, confirming that the morphologies are scale-invariant. 

Next, we turn our attention to the central theme of this paper, viz., the scattering properties of the domain morphology in Fig. \ref{RFIM}. The scattering of a plane wave by a rough surface can yield useful information about the texture of the surface \cite{bs84,wong,wb}. Thus, small-angle scattering experiments (using X-rays, neutrons, etc.) can be used to probe the nature of domain walls separating the components of an inhomogeneous system. These experiments yield the structure factor $S\left(k,\Delta\right)$, which is the Fourier transform of the correlation function. Experimentalists are interested in the large-$k$ (tail) behavior of $S\left(k,\Delta\right)$, which is determined by the small-$r$ (short-distance) behavior of $C\left( r,\Delta\right)$. In the inset of Fig. \ref{SCF}, we plot $1 - C\left( r,\Delta\right)$ vs. $r/\xi$ on a log-log scale. The small-$r$ behavior shows a distinct cusp singularity:  $C\left(r,\Delta\right) \simeq 1 - A\left(r/\xi\right)^{\alpha} + \cdot\cdot\cdot $ with $\alpha \simeq 0.5$. This holds over more than a decade in $r/\xi$-values. A similar cusp has been reported earlier also in the context of {\it fluctuation-dominated phase-separation} (FDPS) \cite{db,mr}. The {\it cusp exponent} $\alpha$ is identical to the {\it roughness exponent} of the domain boundaries. [Notice that the value we obtain for the paramagnetic phase ($\alpha_{\rm para} \simeq 0.5$) differs considerably from the roughness exponent in the ferromagnetic phase. The Middleton-Fisher value for the latter exponent is $\alpha_{\rm ferro} = 0.66 \pm 0.03$, which is consistent with the theoretical result $\alpha_{\rm ferro} = 2/3$ \cite{im76,th90}.] The corresponding interfaces are self-affine fractals with $d_f = d-\alpha$. Therefore, in our present study, $d_f \simeq 2.5$, which is consistent with studies of percolation clusters in the strong-disorder regime of the $d=3$ RFIM by Seppala et al. \cite{spa02} and Ji and Robbins \cite{jr92}.

\begin{center}
\begin{figure}
\includegraphics[width=15cm]{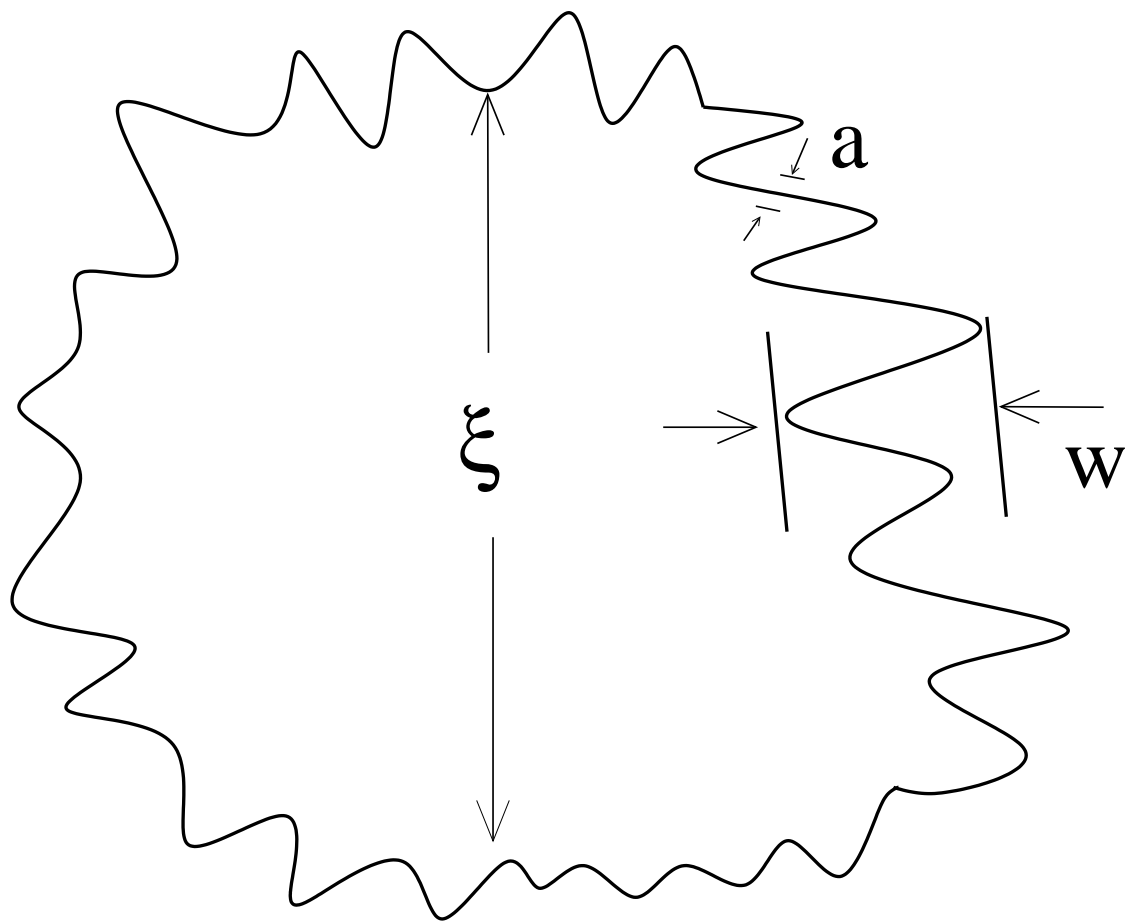}
\caption{Schematic of a domain of size $\xi$. The characteristic interface thickness is $w$, and the microscopic lattice spacing $a=1$.}
\label{DMN}
\end{figure}
\end{center}

In Fig. \ref{DMN}, we show a schematic of a domain of size $\xi$, with an interface of width $w$. There is also a microscopic length scale $a=1$, due to the underlying discreteness of the lattice. The corresponding $C\left(r,\Delta\right)$ would show corrections to scaling, characterized by the parameter $w/\xi$. Systems characterized by a cusp singularity exhibit very rough interfaces with $w\sim\xi$. A novel feature of the present work is that we observe this scattering phenomenology in the paramagnetic phase of the RFIM, i.e., in the absence of interfaces between coexisting equilibrium phases. The lower frames in Fig. \ref{RFIM} show cross-sections of the $d = 3$ snapshots in the upper frames. Notice that the domain structure is fuzzy and subject to large fluctuations, and the boundaries are ill-defined. 

At larger values of $x = r/\xi$, the correlation function is well-approximated as 
\begin{equation}
\label{corrl}
C\left(r,\Delta\right) \simeq 1 - A\left(r/\xi\right)^{\alpha} - B\left(r/\xi\right) + \cdot\cdot\cdot .
\end{equation}
The linear decay in Eq.~(\ref{corrl}) is characteristic of scattering from sharp interfaces in inhomogeneous systems, and is termed the {\it Porod law} \cite{op88}.  With reference to the schematic in Fig. \ref{DMN}, the correlation function $C(r,\Delta)$ exhibits (a) no systematic structure for $r\sim a$; (b) interfacial structure or cusp singularity for $w \gg r \gg a$; (c) Porod decay for $\xi \gg r \gg w.$

The short-distance cusp singularity in $C(r,\Delta)$ has important implications for the structure factor $S\left(k,\Delta\right)$. The scattered intensity now decays with an asymptotic power-law form \cite{bs84,wong,wb} 
\begin{equation}
\label{cusp}
S\left(k,\Delta\right) \sim \tilde{A}\left(\xi k\right)^{-(d+ \alpha)} + \tilde{B} \left(\xi k\right)^{-(d+ 1)},
\end{equation}
valid for $k \ll a^{-1}.$ For $k \sim a^{-1}$, the structure factor becomes flat, corresponding to the absence of structure at microscopic scales. The dominant large-$k$ behavior in Eq.~(\ref{cusp}) is $S(k) \sim \left(\xi k\right)^{-(d+\alpha)}$ with cross-over momentum $k_{c} \sim \xi^{-1}$. 

\begin{center}
\begin{figure}
\includegraphics[width=15cm]{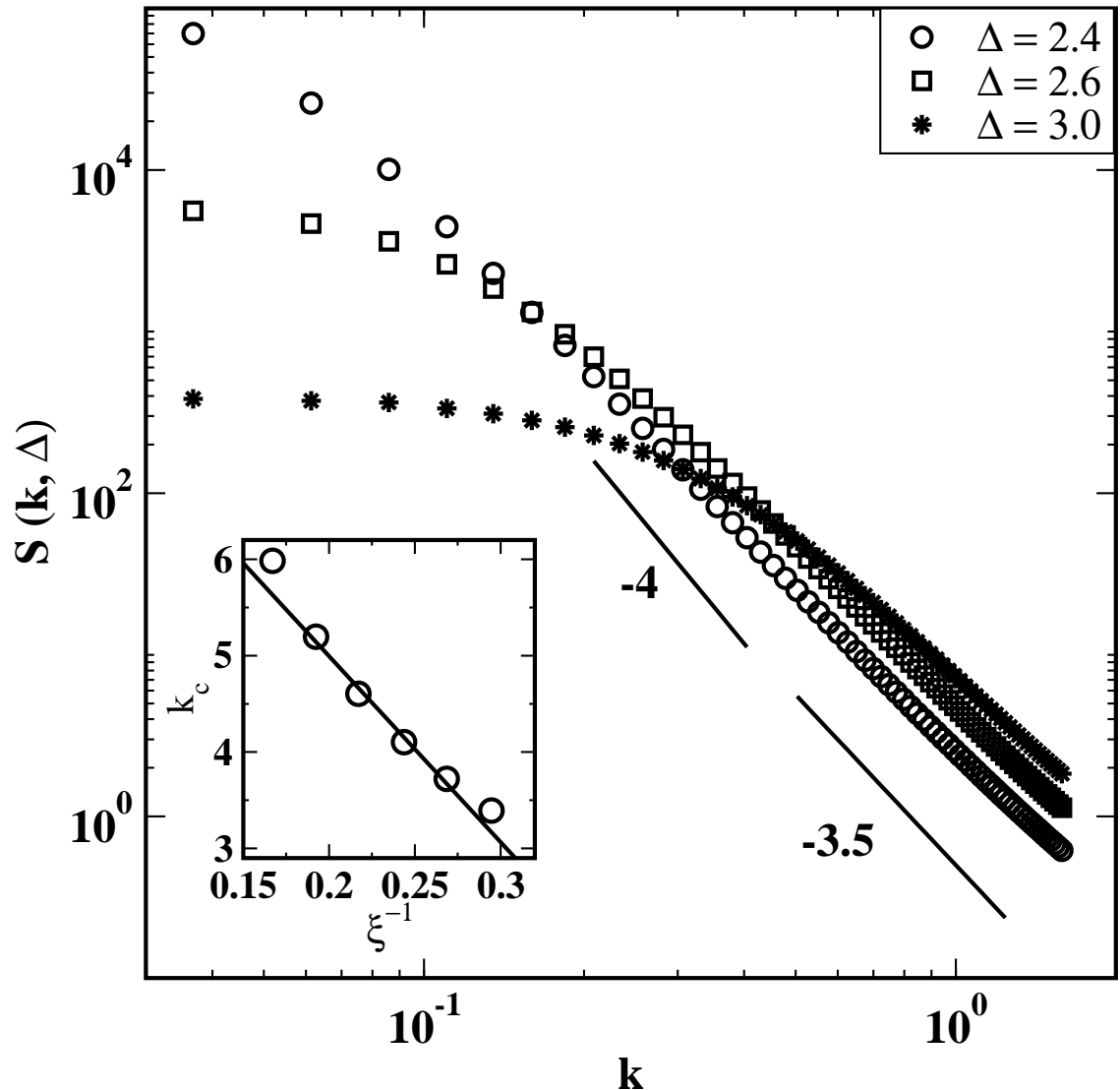}
\caption{Structure factor $\left[S\left(k,\Delta\right) \ \mbox{vs.} \ k\right]$, corresponding to the correlation functions in Fig. 4. The solid lines denotes a {\it Porod regime} $\left[S\left(k,\Delta\right) \sim k^{-4} \right]$ at intermediate $k$-values, which crosses over to an asymptotic {\it cusp regime $\left[S\left(k,\Delta\right) \sim k^{-3.5} \right]$}.  The inset shows the behavior of the cross-over  momentum $k_{c}$ vs. $\xi^{-1}$ for several values of disorder. }
\label{SF}
\end{figure}
\end{center}

In Fig. \ref{SF}, we plot the structure factor for the RFIM with $\Delta > \Delta_{c}$ on a log-log scale.  Our data is consistent with the scaling form in Eq.~(\ref{cusp}). There is a cross-over from a {\it Porod regime} $\left[\mbox{with} \ S\left(k,\Delta\right) \sim k^{-(d+1)}\right]$ at intermediate values of $k$ to an asymptotic {\it cusp regime} $\left[\mbox{with} \ S\left(k,\Delta\right) \sim k^{-(d+\alpha)}, \ \alpha \simeq 0.5\right]$. The inset of Fig. \ref{SF} shows the behavior of the cross-over momentum, which scales as $k_{c} \sim \xi^{-1}$. We conjecture that the crossover is a generic feature in the RFIM as a consequence of  interfacial roughening caused by quenched disorder \cite{huse}. In this context, we consider Refs. \cite{rc93,lc08} where the authors studied domain growth in the $d = 3$ RFIM. They focused on the nonequilibrium evolution of the system after a quench from the paramagnetic phase $\left(\Delta = \infty\right)$ to the ferromagnetic phase $\left(\Delta < \Delta_{c}\right)$. Refs. \cite{rc93,lc08} observe that the scaling functions of the RFIM and the pure Ising system are identical, thereby exhibiting {\it super-universality} \cite{pcp91,bh91,ppr04} or irrelevance of quenched randomness. However, a careful observation of the scaled correlation data (Fig. 2 in \cite{rc93}) for small $r/\xi$ reveals clear deviations from the pure system and so from the Porod law. [Of course, we expect to recover the Porod law in the limit $w/\xi\rightarrow 0.$ This is possible in the domain growth problem as $\xi(t) \rightarrow \infty$ as $t \rightarrow \infty.$]

We conclude this paper with a summary and discussion of our results. We have used a computationally efficient {\it graph-cut method} (GCM) to study the ground-state ($T = 0$) properties of the RFIM in the paramagnetic state. The Boykov-Kolmogorov GCM used by us provides access to the ground-state morphology of large systems. We characterize this morphology using correlation functions and structure factors, which contain information averaged over {\it all} domains and interfaces. The correlation function $C\left(r,\Delta\right)$ is characterized by a universal scaling function for different disorder amplitudes. There are no perceptible {\it corrections to scaling} for different values of $\Delta$, suggesting that the interface thickness $w$ scales with the correlation length $\xi$. At short distances,  $C\left(r,\Delta\right)$ shows a cusp singularity reminiscent of that seen in {\it fluctuation-dominated phase separation} \cite{db,mr}. This is associated with scattering off rough, fractal interfaces. The corresponding structure factor  $S\left(k,\Delta\right)$ shows a crossover from a {\it Porod regime} at intermediate $k$ values, to an asymptotic {\it cusp regime}. These properties should be universal for disordered systems, which are often characterized by rough interfaces. We believe that our results will motivate further analytical and numerical studies of this problem.

GPS and VB would like to acknowledge the support of DST Grant No. SR/S2/CMP-002/2010. We thank S.N. Maheshwari and Chetan Arora for fruitful discussions. We are also grateful to Uma Mudenagudi and Olga Veksler for technical support in programming.

\newpage

\newpage

\newpage

\newpage

\newpage

\newpage

\end{document}